\let\accentvec\vec % necessary to inhibit amsmath warning caused by definition of \vec in llncs
\documentclass[oribibl,envcountsame]{llncs}  % use original bibliography and make theorem-like environments use the same counter
\usepackage{description_logic}
\reporttrue
\usepackage{asymptote}

\spnewtheorem{prop}[theorem]{Proposition}{\bfseries}{\itshape} 
\spnewtheorem{observation}[theorem]{Observation}{\bfseries}{\itshape}

\title{%
  \texorpdfstring{%
    The Complexity of Satisfiability for\\Sub-Boolean Fragments of \ALC%
  }{%
    The Complexity of Satisfiability for Sub-Boolean Fragments of ALC%
  }%
}
\author{Arne Meier\protect\texorpdfstring{\inst{1}}{} and Thomas Schneider\protect\texorpdfstring{\inst{2}}{}}
\institute{%
  Leibniz Universit\"{a}t Hannover, Germany,~
  \email{meier@thi.uni-hannover.de}%
  \and
  University of Manchester, UK,~
  \email{schneider@cs.man.ac.uk}%
}
\authorrunning{A. Meier and T. Schneider}

\begin{document}

\maketitle

\begin{abstract}
  The standard reasoning problem, concept satisfiability, in the basic description logic \ALC
  is \PSPACE-complete, and it is \EXPTIME-complete in the presence of unrestricted axioms.
  Several fragments of \ALC, notably logics in the \FL, \EL, and DL-Lite families,
  have an easier satisfiability problem; sometimes it is even tractable.
  All these fragments restrict the use of Boolean operators in one way or another.
  We look at systematic and more general restrictions
  of the Boolean operators and establish the complexity of the concept satisfiability problem
  in the presence of axioms. We separate tractable from intractable cases.
\end{abstract}

\section{Introduction}

Standard reasoning problems of description logics, such as satisfiability or subsumption,
have been studied extensively. Depending on the expressivity of the logic and the reasoning problem,
the complexity of reasoning for DLs ranging from logics below the basic description logic \ALC to the OWL DL standard \SROIQ
is between tractable and \NEXPTIME.

For \ALC, concept satisfiability is \PSPACE-complete \cite{scsm91} and,
in the presence of unrestricted axioms, it is \EXPTIME-complete due to the correspondence with propositional
dynamic logic \cite{pr78,vawo86,doma00}. Since the standard reasoning tasks are interreducible
in the presence of all Boolean operators, subsumption has the same complexity.

Several fragments of \ALC, such as logics in the \FL, \EL or DL-Lite families,
are well-understood. They often restrict the use of Boolean operators, and it is known that 
their reasoning problems are often easier than for \ALC. 
For instance, concept subsumption with respect to acyclic and cyclic TBoxes, and even with GCIs
is tractable in the logic \EL, which allows only conjunctions and existential restrictions,
\cite{baa03a,bra04}, and it remains tractable under a variety of extensions
such as nominals, concrete domains, role chain inclusions, and domain and range restrictions
\cite{bbl05,bbl08}. However, the presence of universal quantifiers breaks tractability:
Subsumption in \FLzero, which allows only
conjunction and universal restrictions, is \coNP-complete \cite{neb90} and
increases to \PSPACE-complete with respect to cyclic TBoxes \cite{baa96,kn03}
and to \EXPTIME-complete with GCIs \cite{bbl05,hof05}. In \cite{dlnhnm92,dlnn97},
concept satisfiability and subsumption for several logics below and above \ALC that extend \FLzero with
disjunction, negation and existential restrictions and other features,
is shown to be tractable, \NP-complete, \coNP-complete or \PSPACE-complete.
Subsumption in the presence of general axioms is \EXPTIME-complete
in logics containing both existential and universal restrictions plus conjunction or disjunction \cite{gimcwiko02},
as well as in \AL, where only conjunction, universal restrictions and unqualified existential restrictions
are allowed \cite{don03}.
In DL-Lite, where atomic negation, unqualified existential and universal restrictions, conjunctions
and inverse roles are allowed, satisfiability of ontologies is tractable \cite{cgllr05}.
Several extensions of DL-Lite are shown to have tractable and \NP-complete satisfiability
problems in \cite{ackz07,ackz09}.

This paper revisits restrictions to the Boolean operators in \ALC.
Instead of looking at one particular subset of $\{\land,\lor,\neg\}$, 
we are considering all possible sets of Boolean operators, including less commonly used operators
such as the binary exclusive or $\xor$.
Our aim is to find for \emph{every} possible combination of Boolean operators 
whether it makes satisfiability of the corresponding restriction of \ALC hard or easy.
Since each Boolean operator corresponds to a Boolean function---\ie, an $n$-ary function whose arguments and values are in $\{\false,\true\}$---there are infinitely many sets of Boolean operators determining fragments of \ALC.
The complexity of the corresponding concept satisfiability problems without theories, which are equivalent to the satisfiability problems for the corresponding
fragments of multimodal logic, has already been classified in \cite{hescsc08}: it is \PSPACE-complete if
at least the ternary operator $x \land (y \lor z)$ and the constant $\false$ are allowed, \coNP-complete if
at least conjunctions and at most conjunctions plus the constant $\false$ are allowed, and trivial otherwise,
\ie, for all other sets of Boolean operators, every modal formula (concept description) is satisfiable.
We will put this classification into the context of the above listed results for \ALC fragments.

The tool used in \cite{hescsc08} for classifying the infinitely many satisfiability problems was
Post's lattice \cite{pos41},
which consists of all sets of Boolean functions closed under superposition. These sets directly correspond to
all sets of Boolean operators closed under nesting. Similar classifications have been achieved
for satisfiability for classical propositional logic
\cite{le79}, Linear Temporal Logic \cite{bsssv07}, hybrid logic \cite{MMSTWW09}, and for constraint
satisfaction problems \cite{sch78,schnoor07}.

In this paper, we classify the concept satisfiability problems with respect to theories for \ALC fragments obtained by arbitrary sets of Boolean operators. We will separate tractable and intractable cases, showing that these problems are
\begin{itemize}
  \item
    \EXPTIME-hard whenever we allow at least conjunction, disjunction or all self-dual operators,
    where a Boolean function is called self-dual if negating all its arguments negates its value,
  \item
    \PSPACE-hard whenever we allow at least negation or both constants $\false,\true$,
  \item
    \coNP-hard whenever we allow at least the constant $\false$,
  \item
    trivial, which means that all instances are satisfiable, in all other cases.
\end{itemize}
We will also put these results into the context of the above listed results for \ALC fragments.
This is work in progress which we plan to extend by corresponding upper bounds, restricted use of $\exists,\forall$,
and terminological restrictions to TBoxes such as acyclicity and atomic left-hand sides of axioms.
Furthermore, not all results carry over straightforwardly to other reasoning problems because some of the standard
reductions use Boolean operators that are not available in every fragment.

\section{Preliminaries}
\label{sec:prelims}

\paragraph*{Description Logic.}

We use the standard syntax and semantics of \ALC with the Boolean operators $\sqcap$, $\sqcup$, $\neg$, $\top$, $\bot$
replaced by arbitrary operators $o$ that correspond to Boolean functions $f_o$ of arbitrary arity. Let \CONC, \ROLE
and \IND be sets of atomic concepts, roles and individuals. Then the set of \emph{concept descriptions}, for short \emph{concepts}, is defined by
\[
    C := A \mid o(C,\dots,C) \mid \exists R.C \mid \forall R.C,
\]
where $A \in \CONC$, $R \in \ROLE$, and $o$ is a Boolean operator. 
A \emph{general concept inclusion (GCI)} is an axiom of the form $C \sqsubseteq D$
where $C,D$ are concepts. We use ``$C \equiv D$'' as the usual syntactic sugar for ``$C \sqsubseteq D$ and $D \sqsubseteq C$''.
A \emph{TBox} is a finite set of GCIs without restrictions.
An \emph{ABox} is a finite set of axioms of the form $C(x)$ or $R(x,y)$,
where $C$ is a concept, $R \in \ROLE$ and $x,y \in \IND$.
An \emph{ontology} is the union of a TBox and an ABox. This simplified view suffices for our purposes.

An \emph{interpretation} is a pair $\calI = (\Delta^\calI, \cdot^\calI)$, where $\Delta^\calI$ is a nonempty set and $\cdot^\calI$ is a mapping from $\CONC$ to $\mathfrak{P}(\Delta^\calI)$, from $\ROLE$ to
$\mathfrak{P}(\Delta^\calI \times \Delta^\calI)$ and from $\IND$ to $\Delta^\calI$
that is extended to arbitrary concepts as follows:
\begin{align*}
%   o(C_1,\dots,C_n)^\calI &= \Big\{x \in \Delta^\calI ~\Big|~ f_o\Big(\|x \in C_1^\calI\|, \dots, \|x \in C_n^\calI\|\Big) = 1\Big\} \\
  o(C_1,\dots,C_n)^\calI &= \{x \in \Delta^\calI \mid f_o(\|x \in C_1^\calI\|, \dots, \|x \in C_n^\calI\|) = \true\}, \\
                         & \qquad\text{where~} \|x \in C_1^\calI\|=\true \text{~if~} x \in C_1^\calI
                                 \text{~and~} \|x \in C_1^\calI\|=\false \text{~if~} x \notin C_1^\calI,    \\
  \exists R.C^\calI      &= \{x\in\Delta^\calI \mid \{y\in C^\calI \mid (x,y)\in R^\calI\} \neq \emptyset\},\\
  \forall R.C^\calI      &= \{x\in\Delta^\calI \mid \{y\in C^\calI \mid (x,y)\notin R^\calI\} = \emptyset\}.
\end{align*}
An interpretation \calI \emph{satisfies} the axiom $C \sqsubseteq D$, written $\calI \models C \sqsubseteq D$, if
$C^\calI \subseteq D^\calI$. Furthermore, $\calI$ satisfies $C(x)$ or $R(x,y)$ if $x^\calI \in C^\calI$
or $(x^\calI,y^\calI) \in R^\calI$. An interpretation \calI satisfies a TBox (ABox, ontology)
if it satisfies every axiom therein. It is then called a \emph{model} of this set of axioms.

Let $B$ be a finite set of Boolean operators and use $\Conc(B)$ and $\Ax(B)$ to denote the set of all concepts 
and axioms using only operators in $B$.
The following decision problems are of interest for this paper.
\begin{description}
  \itemsep4pt
  \item[\textbf{Concept satisfiability $\ALCCSAT(B)$}:]~\\
    Given a concept $C \in \Conc(B)$, is there an interpretation \calI \st $C^\calI\neq\emptyset$\,?
  \item[\textbf{TBox satisfiability $\ALCTSAT(B)$}:]~\\
    Given a TBox $\calT \subseteq \Ax(B)$, is there an interpretation \calI \st $\calI \models \calT$\,?
  \item[\textbf{TBox-concept satisfiability $\ALCTCSAT(B)$}:]~\\
    Given $\calT \subseteq \Ax(B)$ and $C \in \Conc(B)$, is there an \calI \st $\calI \models \calT$
    and $C^\calI\neq\emptyset$\,?
  \item[\textbf{Ontology satisfiability $\ALCOSAT(B)$}:]~\\
    Given an ontology $\calO \subseteq \Ax(B)$, is there an interpretation \calI \st $\calI \models \calO$\,?
  \item[\textbf{Ontology-concept satisfiability $\ALCOCSAT(B)$}:]~\\
    Given $\calO \subseteq \Ax(B)$ and $C \in \Conc(B)$, is there an \calI \st $\calI \models \calO$
    and $C^\calI\neq\emptyset$\,?
\end{description}
These problems are interreducible independently of $B$ in the following way:
\begin{align*}
  & \ALCCSAT(B) \leqlogm \ALCOSAT(B) \\
  & \ALCTSAT(B) \leqlogm \ALCTCSAT(B) \leqlogm \ALCOSAT(B) \equivlogm \ALCOCSAT(B)
\end{align*}
The reasons are: a concept $C$ is satisfiable iff the ontology $\{a:C\}$ is satisfiable,
for some individual $a$;~ a terminology $\calT$ is satisfiable iff a fresh atomic concept $A$ is satisfiable w.r.t. $\calT$; $C$ is satisfiable w.r.t.\ $\calT$ iff $\calT \cup \{a:C\}$ is satisfiable, for a fresh individual $a$.

% \problemdef{$\ALCCSAT(B)$}{A concept $C$ using the Boolean connectives in $B$ only.}{Yes iff there exists an interpretation \calI \st $C^\calI\neq\emptyset$}
% 
% \problemdef{$\ALCOCSAT(B)$}{A concept $C$ and an ontology \calO, which both use the Boolean connectives in $B$ only.}{Yes iff there exists an interpretation \calI \st $\calI\models\calO$ and $C^\calI\neq\emptyset$}
% 
% \problemdef{$\ALCOSAT(B)$}{An ontology \calO using the Boolean connectives in $B$ only.}{Yes iff there exists an interpretation \calI \st $\calI\models\calO$}
% 
% \problemdef{$\MSAT(B)$}{A modal logic formula $\varphi$ using the Boolean connectives in $B$ only.}{Yes iff $\varphi$ is satisfiable.}

\paragraph*{Complexity Theory.}

We assume familiarity with the standard notions of complexity theory as, e.\,g., defined in \cite{pap94}.
In particular, we will make use of the classes $\P$, $\NP$, $\coNP$, $\PSPACE$, and $\EXPTIME$,
as well as logspace reductions $\leqlogm$.

\paragraph*{Boolean operators.}
  This study aims at being complete with respect to Boolean operators, which correspond to Boolean functions.
  A set of Boolean functions is called a \emph{clone} if it is closed under superpositions of functions,
  \ie, nesting of operators. The lattice of all clones has been established in \cite{pos41},
  see \cite{bocrrevo03} for a more succinct but complete presentation. Via the inclusion structure,
  lower and upper complexity bounds carry over to higher and lower clones.
  We will therefore only state our results for minimal and maximal clones.

  Given a finite set $B$ of functions, the smallest clone containing $B$ is denoted by $[B]$.
  The set $B$ is called a \emph{base} of $[B]$, but $[B]$ often has other bases as well.
  On the operator side, $[B]$ consists of all operators obtained by nesting operators
  in $B$ into each other. For example, nesting of binary conjunction yields conjunctions
  of arbitrary arity. The table below lists all clones that we will refer to, using
  the following definitions.
  A Boolean function $f$ is called \emph{self-dual} if
  $f(\overline{x_1},\dots,\overline{x_n}) = \overline{f(x_1,\dots,x_n)}$,
  \emph{$c$-reproducing} if $f(c,\dots,c) = c$, and \emph{$c$-separating} if there is an $1\leq i\leq n$ \st for each $(b_1,\dots,b_n)\in f^{-1}(c)$ $b_i=c$ for $c\in\{\true,\false\}$. 
  The symbol $\xor$ denotes the binary exclusive or.

  \begin{center}
    \begin{small}
      \begin{tabular}{@{~~~}l@{~~~~~}l@{~~~~~}l@{~~~}}
        \hline\rule{0pt}{8pt}%
        Clone          & Description                      & Base                     \\
        \hline\rule{0pt}{8pt}%
        $\CloneBF$     & all Boolean functions            & $\{\land, \neg\}$ \\
        $\CloneM$      & All monotone functions & $\{\land, \lor, \false, \true\}$ \\
%        \ifreport
%          $\CloneS_1$    &                                  & $\{x \land \overline{y}\}$ \\
%        \fi
        $\CloneS_{11}$ &  \true-separating, monotone function& $\{x \land (y \lor z),\false\}$ \\
        $\CloneD$      & self-dual functions              & $\{ (x \land \overline{y}) \lor (x \land \overline{z}) \lor (\overline{y} \land \overline{z}) \}$ \\
        $\CloneE$      & conjunctions and constants       & $\{\land,\false,\true\}$ \\
        $\CloneE_0$    & conjunctions and $\false$        & $\{\land,\false\}$   \\
        $\CloneV_0$    & disjunctions and $\false$        & $\{\lor,\false\}$    \\
        $\CloneR_1$    & $\true$-reproducing functions          & $\{\lor,x \xor y \xor \true\}$ \\
        $\CloneR_0$    & $\false$-reproducing functions          & $\{\land,\xor\}$ \\
        $\CloneN_2$    & negation                         & $\{\neg\}$ \\
        $\CloneI$      & identity functions and constants & $\{\operatorname{id},\false,\true\}$ \\
        $\CloneI_0$    & identity functions and $\false$  & $\{\operatorname{id},\false\}$ \\
        \hline
      \end{tabular}
    \end{small}
  \end{center}
\ifreport
  The following lemma will help restrict the length
  of concepts in some of our reductions. It shows that for
  certain sets $B$, there are always short concepts representing the
  functions $\land$, $\lor$, or $\lnot$, respectively.  Points (2) and (3)
  follow directly from the proofs in \cite{le79}, Point (1) is
  Lemma~1.4.5 from \cite{schnoor07}.

\begin{lemma}\label{lem:lewis-schnoor}
  Let $B$ be a finite set of Boolean functions.
  \begin{enumerate}
  \item
    If $\CloneV\subseteq[B]\subseteq\CloneM$
    ($\CloneE\subseteq[B]\subseteq\CloneM$, resp.), then there exists a $B$-formula $f(x,y)$ such that $f$ represents
    $x\vee y$ ($x\wedge y$, resp.) and each of the variables $x$ and $y$ occurs exactly once in $f(x,y)$.
  \item
    If $[B]=\CloneBF$, then there are $B$-formulae $f(x,y)$ and
    $g(x,y)$ such that $f$ represents $x\vee y$, $g$ represents $x\wedge y$, and both variables occur in each of these
    formulae exactly once.
  \item
    If $\CloneN\subseteq[B]$, then there is a $B$-formula
    $f(x)$ such that $f$ represents $\neg x$ and the variable $x$ occurs in $f$ only once.
  \end{enumerate}
\end{lemma}
\fi

% \begin{figure}[ht]
% \centering
% 
%\begin{asy}
%   import lattice;
%   defaultpen(0.8+fontsize(9));
%   Lattice lattice = Lattice(0.9cm, 0.65cm, 0.28cm);
% 
%   lattice.draw();
%\end{asy}
% 
% \caption{Post's Lattice.}
% \label{fig:lattice}
% 
% \end{figure}

\paragraph*{Auxiliary results.}
The following lemmata contain technical results that will be useful to formulate our main results.
We use $\STARSAT(B)$ to speak about any of the four satisfiability problems $\ALCTSAT,\ALCTCSAT,\ALCOSAT$ and $\ALCOCSAT$ introduced above.

\begin{lemma}\label{lemma:topbot-always-above-neg}
	Let $B$ be a finite set of Boolean functions. If $\CloneN_2\subseteq[B]$, then it holds that $\STARSAT(B)\equivlogm\STARSAT(B\cup\{\true,\false\})$.
\end{lemma}
\begin{Proof}
  It is easy to observe that the concepts $\true$ and $\false$ can be simulated by fresh atomic concepts
  $T$ and $B$, using the axioms $\lnot T\dsub T$ and $B\dsub\lnot B$. 
\end{Proof}

\begin{lemma}\label{lem:TCSAT_reduces_to_TSAT_with_true}
Let $B$ be a finite set of Boolean functions. Then it holds that $\ALCTCSAT(B)\leqlogm\ALCTSAT(B\cup\{\true\})$.
\end{lemma}
\begin{Proof}
It can be easily shown that $\encoding{C,\calT}\in\ALCTCSAT(B)$ iff $\encoding{\calT\cup\{\true\dsub\exists R.C\}}\in\ALCTSAT(B\cup\{\true\})$, where $R$ is a fresh relational symbol. For "$\Rightarrow$" observe that for the satisfying interpretation $\calI=(\Delta^\calI,\cdot^\calI)$ there  must be a world $w'$ where $C$ holds and then from every world $w\in\Delta^\calI$ there can be an $R$-edge from $w$ to $w'$ to satisfy $\calT\cup\{\true\dsub\exists R.C\}$. For "$\Leftarrow$" note that for a satisfying interpretation $\calI=(\Delta^\calI,\cdot^\calI)$ all axioms in $\calT\cup\{\true\dsub\exists R.C\}$ are satisfied. In particular the axiom $\true\dsub\exists R.C$. Hence there must be at least one world $w'$ \st $w'\models C$. Thus $\calI\models\calT$ and $C^\calI\supseteq\{w'\}\neq\emptyset$.
\end{Proof}

\medskip\noindent
Furthermore, we observe that, for each set $B$ of Boolean functions with $\true,\false\in[B]$, we can simulate the negation of an atomic concept using a fresh atomic concept $A$ and role $R_A$: if we add the axioms $A\equiv\exists R_A.\true$ and $A'\equiv\forall R_A.\false$ to the given terminology $\calT$, then each model of \calT has to interpret $A'$ as the complement of $A$.

\section{Complexity results for \CSAT}

The following classification of concept satisfiability has been obtained in \cite{hescsc08}.

% \section{Concept Satisfiability $\ALCCSAT$}
% \begin{theorem}[\cite{hescsc08}]\label{thm:ALCCSAT_results}Let $B$ be a finite set of Boolean functions.
%         \begin{enumerate}
%                 \item If $\{x\land(y\lor z),\false\}\subseteq B$, then $\ALCCSAT(B)$ is \PSPACE-complete.
%                 \item If $\{\land,\false\}\subseteq B$, then $\ALCCSAT(B)$ is \co\NP-complete.
%                 \item If $\{\lor,\equiv\}\subseteq B$, then $\ALCCSAT(B)$ is trivial.
%                 \item Otherwise $\ALCCSAT(B)\in\P$.
%         \end{enumerate}
% \end{theorem}
\begin{theorem}[\cite{hescsc08}]Let $B$ be a finite set of Boolean functions.
	\begin{enumerate}
		\item If $\CloneS_{11}\subseteq[B]$, then $\ALCCSAT(B)$ is \PSPACE-complete.
		\item If $[B]\in\{\CloneE,\CloneE_0\}$, then $\ALCCSAT(B)$ is \co\NP-complete.
		\item If $[B]\subseteq\CloneR_1$, then $\ALCCSAT(B)$ is trivial.
		\item Otherwise $\ALCCSAT(B)\in\P$.
	\end{enumerate}
\end{theorem}
% \begin{Proof}
% This follows directly by Theorem \ref{thm:ALCCSAT_results} in combination with Lemmas \ref{lem:lewis-schnoor} and \ref{lem:baseIndependence}.
% \end{Proof}

Part (1) is in contrast with the \coNP-completeness of \ALU satisfiability \cite{scsm91} 
because the operators in \ALU can express the canonical base of $\CloneS_{11}$.
The difference is caused by the fact that \ALU allows only \emph{unqualified} 
existential restrictions. Part (2) generalises the \coNP-completeness of \ALE 
satisfiability, where hardness is proven in \cite{dlnhnm92} without using atomic negation.
It is in contrast with the tractability of \AL satisfiability \cite{dlnn97}, again because
of the unqualified restrictions. Part (3) generalises the known fact that every
\EL, \FLzero, and $\FL^{-}$ concept is satisfiable. The results for logics in the
DL-Lite family cannot be put into this context because DL-Lite quantifiers are unqualified.

\section{Complexity Results for \TSAT, \TCSAT, \OSAT, \OCSAT}
In this section we will completely classify the above mentioned satisfiability problems for their tractability with respect to sub-Boolean fragments and put them into context with existing results
for fragments of \ALC.

\paragraph*{Main results.}
Due to the interreducibilities stated in Section \ref{sec:prelims}, it suffices to 
show lower bounds for \ALCTSAT\ and upper bounds for \ALCOCSAT.

\begin{theorem}\label{thm:ALCOCSAT_results}
  Let $B$ be a finite set of Boolean functions.
	\begin{enumerate}
		\item If $\land\in B$ or $\lor\in B$, then $\ALCTCSAT(B)$ is \EXPTIME-hard.\\ If also $\top\in B$, then even $\ALCTSAT(B)$ is \EXPTIME-hard.
                \item If all functions in $B$ are self-dual, then $\ALCTSAT(B)$ is \EXPTIME-hard.
		\item If $\lnot\in B$ or $\{\true,\false\}\subseteq B$, then $\ALCTSAT(B)$ is \PSPACE-hard.
		\item If all functions in $B$ are $\false$-reproducing, then $\ALCTSAT(B)$ is trivial.\\[-.3em]
		\item If $\false\in B$, then $\ALCTCSAT(B)$ is \coNP-hard.
		\item If all functions in $B$ are $\true$-reproducing, then $\ALCOCSAT(B)$ is trivial.
	\end{enumerate}
\end{theorem}
\begin{Proof}
% (1)--(3) will be proven through Lemma \ref{lem:ALCOCSAT_intractable}, and Lemma \ref{lem:ALCOCSAT_R1_trivial} will prove (4).
Parts 1.--6. are formulated as Lemmas 
\ref{lem:ALCOCSAT_EXPTIME-hard},
\ref{lem:ALCOSAT_I_PSPACE_HARD},
\ref{lem:ALCOCSAT_N2_PSPACE-HARD},
\ref{lem:ALCOCSAT_R1_trivial},
\ref{lem:ALCTSAT_R0_trivial},
\ref{lem:ALCOCSAT_I0_CONP-HARD},
and are proven below. The second part of (1.) follows from Lemma \ref{lem:ALCOCSAT_EXPTIME-hard} in combination with Lemma \ref{lem:TCSAT_reduces_to_TSAT_with_true}.
\end{Proof}

\ifreport
In order to generalize these results, we need to prove the following lemma. It states the base independence that will lead
to the more general results in Corollary \ref{cor:ALCOCSAT_results_full} and Corollary \ref{cor:ALCTSAT_results_full}.
\begin{lemma}\label{lem:Base_Independence}
Let $B_1,B_2$ be two sets of Boolean functions \st $[B_1]=[B_2]$. Then $\STARSAT(B_1)\leqlogm\STARSAT(B_2)$.
\end{lemma}
\begin{Proof}
According to \cite[Theorem 3.6]{hescsc08}, we translate for any given instance each Boolean formula (hence each side of an axiom) into a Boolean circuit over the basis $B_1$. This circuit can be easily transformed into a circuit over the basis $B_2$. This new circuit will be expressed by several new axioms that are constructed in the style of the formulae in \cite{hescsc08}:
\begin{itemize}
	\item For input gates $g$, we add the axiom $g\equiv x_i$.
	\item If $g$ is a gate computing the Boolean function $\phi$ and $h_1,\dots,h_n$ are the respective predecessor gates in this circuit, we add the axiom $g\equiv\phi(h_1,\dots,h_n)$.
	\item For $\exists R$-gates $g$, we add the axiom $g\equiv \exists R.h$.
        \item Analogously for $\forall R$-gates.
\end{itemize}

For each axiom $A\dsub B$, let $g_{out}^A$ and $g_{out}^B$ be the output gates of the appropriate circuits. Then we need to add one new axiom $g_{out}^A\dsub g_{out}^B$ to ensure the axiomatic property of $A\dsub B$. If the translated formula $\psi$ is a given concept expression (relevant for the problems $\ALCTCSAT,\ALCOCSAT$), the translated concept is mapped to the respective out-gate $g_{out}^\psi$.

This reduction is computable in logarithmic space and its correctness can be shown in the same way as in the proof of Theorem 3.6 in \cite{hescsc08}.
\end{Proof}

\fi

As a consequence of Theorem \ref{thm:ALCOCSAT_results} in combination with 
\ifreport
Lemma \ref{lem:Base_Independence}
\else
Lemma 6 in \cite{mesc10}
\fi
, we obtain the following two corollaries that generalise the results to arbitrary bases for all four satisfiability problems.

\begin{corollary}\label{cor:ALCOCSAT_results_full}
  Let $B$ be a finite set of Boolean functions and $\STARSAT'$ one of the problems 
  $\TCSAT$, $\OSAT$ and $\OCSAT$.
	\begin{enumerate}
		\item If $\CloneE_0\subseteq[B]$ or $\CloneV_0\subseteq[B]$, and $[B]\subseteq\CloneM$, then $\STARSAT'(B)$ is \EXPTIME-hard.
    \item If $[B]=\CloneD$ or $[B]=\CloneBF$, then $\STARSAT'(B)$ is \EXPTIME-hard.
		\item If $\CloneN_2\subseteq[B]$ or $\CloneI\subseteq[B]$, then $\STARSAT'(B)$ is \PSPACE-hard.
		\item If $[B]=\CloneI_0$, then $\STARSAT'(B)$ is \coNP-hard.
		\item If $[B]\subseteq\CloneR_1$, then $\STARSAT'(B)$ is trivial.
	\end{enumerate}
\end{corollary}
\begin{corollary}\label{cor:ALCTSAT_results_full}
  Let $B$ be a finite set of Boolean functions.
	\begin{enumerate}
		\item If $\CloneE\subseteq[B]$ or $\CloneV\subseteq[B]$, and $[B]\subseteq\CloneM$, then $\ALCTSAT(B)$ is \EXPTIME-hard.
    \item If $[B]=\CloneD$ or $[B]=\CloneBF$, then $\ALCTSAT(B)$ is \EXPTIME-hard.
		\item If $\CloneN_2\subseteq[B]$ or $\CloneI\subseteq[B]$, then $\ALCTSAT(B)$ is \PSPACE-hard.
		\item If $[B]\subseteq\CloneR_0$, or $[B]\subseteq\CloneR_1$, then $\TSAT(B)$ is trivial.
	\end{enumerate}
\end{corollary}

Part (1) generalises the \EXPTIME-hardness of subsumption for \FLzero and \AL with respect to
GCIs \cite{gimcwiko02,don03,hof05}. It is in contrast to the tractability of
subsumption with respect to GCIs in \EL because our result does not separate the two types of
restriction, because \EL has only existential restriction, and our results do not (yet) consider existential, resp., universal restrictions separately. This undermines the observation that, for negation-free fragments, the choice of
the quantifier affects tractability and not the choice between conjunction and disjunction.
Again, DL-Lite cannot be put into this context because of the unqualified restrictions.

Parts (2)--(4) (resp. (2) and (3) for Corollary \ref{cor:ALCTSAT_results_full}) show that satisfiability with respect to theories is already intractable for even
smaller sets of Boolean operators. One reason is that sets of axioms already contain limited forms
of implication and conjunction. This also causes the results of this analysis to differ from similar analyses
for related logics in that hardness already holds for bases of clones that are comparatively
low in Post's lattice.

Due to Post's lattice, our analysis is complete for dividing the fragments into tractable and intractable cases.

\subsection*{Proofs of the main results.}

\begin{lemma}\label{lem:ALCOCSAT_R1_trivial}
  Let $B$ be a finite set of Boolean functions \st $B$ contains only $\true$-reproducing functions.
  Then $\ALCOCSAT(B)$ is trivial.
\end{lemma}

\begin{Proof}
According to Post's lattice, every $B$ that does not fall under Theorem \ref{thm:ALCOCSAT_results}\,(1)--(4)+(6)
contains only $\true$-reproducing functions.
Hence the following interpretation satisfies any instance $(\calO,C)$:~
$\calI=(\{w\},\cdot^\calI)$ \st $A^\calI=\{w\}$ for each atomic concept $A$, 
$r^\calI=\{(w,w)\}$ for each role $r$, and $a^\calI=w$ for each individual $a$.
It then holds trivially that $\calI\models\calO$ and $C^\calI=\{w\}\neq\emptyset$.
\end{Proof}

\begin{lemma}\label{lem:ALCTSAT_R0_trivial}
	  Let $B$ be a finite set of Boolean functions \st $B$ contains only $\false$-reproducing functions. Then $\TSAT(B)$ is trivial.
\end{lemma}
\begin{Proof}
	The intepretation $\calI=(\{w\},\cdot^\calI)$ with $A^\calI=\emptyset$ for each atomic concept $A$, and $r^\calI=\{(w,w)\}$ for each role $r$ satisfies any instance $\calT$ for $\TSAT(B)$, where $B$ contains only $\false$-reproducing functions. This follows from the observation that for each axiom $A\dsub B$ in $\calT$ both sides are always falsified by $\calI$ (because every atomic concept is falsified, and we only have $\false$-reproducing operators as connectives). This can be shown by an easy induction on the concept  structure. Please note that we need to construct a looping node concerning the transition relations due to the fact that we need to falsify axioms with $\forall r.\false$ on the left side for some relation $r$. If we set $r^\calI=\emptyset$ then this expression would be satisfied and would contradict our argumentation for the axiom $\forall r.\false\dsub\false$. Moreover this construction cannot fulfill wrongly the left side of an axiom because of the absence of $\top$ and as no atomic concept has instances with $w$.
\end{Proof}

% \subsection{Intractable Cases}
% \begin{lemma}Let $B$ be a finite set of Boolean functions.
% 	\begin{enumerate}
% 		\item If $\land\in B$, $\lor\in B$, or self-dual functions can be expressed in $B$, then $\ALCTSAT(B)$ is \EXPTIME-hard.
% 		\item If $\lnot\in B$ or $\{\true,\false\}\subseteq B$, then $\ALCTSAT(B)$ is \PSPACE-hard.
% 		\item If $\false\in B$, then $\ALCTSAT(B)$ is \coNP-hard.
% 	\end{enumerate}
% 	\label{lem:ALCOCSAT_intractable}
% \end{lemma}
\begin{lemma}\label{lem:ALCOCSAT_EXPTIME-hard}
  Let $B$ be a finite set of Boolean functions with $\land\in B$, or $\lor\in B$. Then $\ALCTCSAT(B)$ is \EXPTIME-hard. If all self-dual functions can be expressed in $B$, then $\ALCTSAT(B)$ is \EXPTIME-hard.
\end{lemma}
\begin{Proof}
The cases $\land\in B$ and $\lor\in B$ follow from \cite{gimcwiko02}.
  The remaining case for the self-dual functions follows from
  \ifreport
    Lemmas \ref{lem:lewis-schnoor} and \ref{lemma:topbot-always-above-neg},
  \else
    Lemma \ref{lemma:topbot-always-above-neg},
  \fi
  as all self-dual functions in combination with the constants $\true,\false$ (to which we have access as $\lnot$ is self-dual) can express any arbitrary Boolean function.
\end{Proof}

\begin{lemma}\label{lem:ALCOSAT_I_PSPACE_HARD}
Let $B$ be a finite set of Boolean functions \st $\{\false,\true\}\subseteq B$. Then $\ALCTSAT(B)$ is \PSPACE-hard.
\end{lemma}
\begin{Proof}
To prove \PSPACE-hardness, we state a $\leqcd$-reduction from $\QBFthreeSAT$ to $\ALCTSAT(B)$ and only allow $\false$ and $\true$ as available functions in $B$. Let $\varphi\equiv \Game_1 x_1\Game_2 x_2 \cdots \Game_n x_n (C_1\land\cdots\land C_m)$ be a quantified Boolean formula and $\Game_i\in\{\exists,\forall\}$. 
In the following we construct a TBox $\calT\subseteq\Ax(B)$ \st $\varphi\equiv\true$ if and only if $\calT\in\ALCTSAT(B)$, where $B$ consists only of $\true$ and $\false$.

We are first adding the following axioms to the TBox $\calT$ using atomic concepts $d_0,\dots,d_n,x_1,\dots,x_n,x_1',\dots,x_n'$ and roles $R_r,R_1,\dots,R_n,S,R_{x_1},\dots,R_{x_n},$ $R_{d_1},\dots,R_{d_n},$ $R_{C_1},\dots,R_{C_m},P_{11},P_{21},P_{31},\dots,P_{1m},P_{2m},P_{3m}$.
The atomic concepts $d_i$ stand for levels, $x_i$ and $x_i'$ for assigning truth values to the variables.
\begin{align}
\nonumber&\text{Initial starting point:}\\\label{eq:init}
&\,\{\true\dsub\exists S.d_0\}\\
\nonumber&\text{$x_i$ is the negation of $x_i'$:}\\\label{eq:varNeg}
&\set{x_i\equiv \exists R_{x_i}.\true}{1\leq i\leq n}\cup\set{x_i'\equiv \forall R_{x_i}.\false}{1\leq i\leq n}\\
\nonumber&\text{in each level $d_i$ we have $R_{i+1}$-successors where $x_{i+1}$ and $x_{i+1}'$ hold:}\\\label{eq:varBranch}
&\set{d_i\dsub \exists R_{{i+1}}. x_{i+1}}{0\leq i <n}\cup\set{d_i\dsub \exists R_{{i+1}}. x'_{i+1}}{0\leq i <n}\\
\nonumber&\text{the levels $d_i$ are disjoint and we have succeeding levels:}\\\label{eq:disjunctLevels}
&\set{d_i\dsub\forall R_{i+1}.d_{i+1}}{0\leq i < n}\cup\nonumber\\ &\quad \set{d_i\dsub\exists R_{d_i}.\true,d_j\dsub\forall R_{d_i}.\false}{0\leq i<j \leq n}\\
\nonumber&\text{$x_i$ and $x_i'$ carry over:}\\\label{eq:varCarry}
&\set{x_i \dsub \forall R_{j}.x_i}{1\leq i<j \leq n}\cup\set{x_i' \dsub \forall R_{j}.x_i'}{1\leq i<j \leq n}
%\\
%\nonumber&\text{Preventing new trees from starting within a tree:}\\\label{eq:prevent}
%&\set{x_i\dsub \exists R_{x_r}.\true,~~ x_i'\dsub\exists R_{x_r}.\true}{1\leq i\leq n}\cup\{x_r\dsub\forall R_{x_r}.\false\}
\end{align}

Now \calT is consistent, and each of its models contains a tree-like substructure similar to the one depicted in Figure \ref{fig:OSAT_I_PSPACE_HARDNESS_PROOF}.
The \emph{root} of this substructure is an instance of $d_0$. The individuals at depth $n$ counting from the root
are called \emph{leaves}.

\begin{wrapfigure}{r}{0.25\textwidth}
	\centering
 \includegraphics{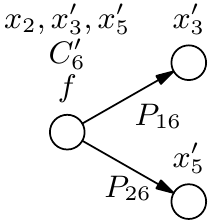}
%\begin{asy}
% 	import circle;
% 	fsize = fontsize(9pt);
% 	defaultpen(fsize);
% 	dy=20;
% 	dx=35;
% 	Kreis s1 = Kreis(0,0,r);
% 	string[] labels1 = {"$x_2,x_3',x_5'$","$C_6'$","$f$"};
% 	s1.labelN(labels1);
% 	s1.draw();
% 	
% 	Kreis s2 = Kreis(dx,dy,r);
% 	string[] labels2 = {"$x_3'$"};
% 	s2.labelN(labels2);
% 	s2.draw();
% 
% 	Kreis s3 = Kreis(dx,-dy,r);
% 	string[] labels3 = {"$x_5'$"};
% 	s3.labelN(labels3);
% 	s3.draw();
% 	kante(s1,s2,"$P_{16}$",SE);
% 	kante(s1,s3,"$P_{26}$",S);
%\end{asy}
	\caption{clause $C_6\equiv \overline x_2\lor x_3\lor x_5$}
	\label{fig:OSAT_I_PSPACE_HARDNESS_PROOF_CRUX}
	\vspace{-.5cm}
\end{wrapfigure}

Please note that each individual in $\Delta^\calI$ is an instance of either $x_i$ or $x_i'$ because of axiom \axref{eq:varNeg}. In particular, this holds for the leaves. Furthermore, this enforcement does not contradict the level-based labeling of the $x_i$---\eg, the atomic concepts $x_i$ and $x_i'$ ``labeled in $d_0$'' are \emph{not} carried forward to the next levels because axiom \axref{eq:varCarry} states this carry only if $j> i$.

In the remaining part, we need to ensure the following, where $C_j$ is an arbitrary clause in $\varphi$. Each leaf $w$ is an instance of the atomic concept $C_j$ if and only if the combination of the $x_i$-values in $w$ satisfies the clause $C_j$. In order to achieve this, we again use two complementary atomic propositions $C_j$ and $C_j'$. The $C_j'$ must be enforced in all leaves where \emph{all} literals of $C_j$ are set to false. For a literal $\ell\in\{x_1,\overline x_1,\dots,x_n,\overline x_n\}$, use $\widetilde \ell$ to denote the atomic concept $x_i$ if $\ell=\overline x_i$ and $x_i'$ if $\ell=x_i$. The correct labeling of the leaves by the $C_j$ and $C_j'$ is ensured by adding the following axioms to $\calT$, which enforce substructures as depicted for the example in Figure \ref{fig:OSAT_I_PSPACE_HARDNESS_PROOF_CRUX}:
\begin{align}
	&\left\{\widetilde l_{1j}\dsub \exists P_{1j}.\true,~~
                \widetilde l_{2j}\dsub\forall P_{1j}.\widetilde l_{2j},~~
                \exists P_{1j}.\widetilde l_{2j}\dsub\exists P_{2j}.\true, \right.\nonumber\\
	&\left.\left.\qquad
                \widetilde l_{3j}\dsub\forall P_{2j}.\widetilde l_{3j},~~
                \exists P_{2j}.\widetilde l_{3j}\dsub C_j',
               \;\right|\;
               C_j= l_{1j}\lor l_{2j}\lor l_{3j} \text{ in } \varphi\right\}~\cup\label{eq:invokeClause}\displaybreak[1]\\
	&\set{C_j'\dsub f}{1\leq j\leq m}\cup\label{eq:labelF}\\
	&\,\{f\dsub \exists F.\true,~~ f'\dsub \forall F.\false\}~\cup\\
	&\set{C_j\equiv\exists R_{C_j}.\true,~~ C_j'\equiv\forall R_{C_j}.\false}{1\leq j\leq m}\label{eq:clauseNeg}
\end{align}

	\begin{wrapfigure}[32]{r}{0.42\textwidth}
		\centering\vspace{-0.5cm}
		
                \includegraphics{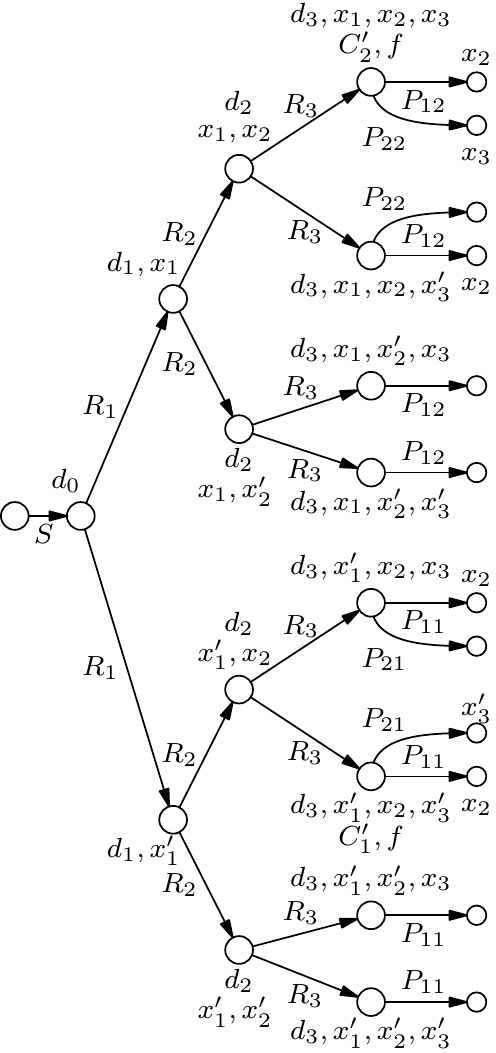}
		\caption{Essential part of the interpretation for the qBf $\varphi=\exists x_1\forall x_2\exists x_3 (x_1\lor\lnot x_2\lor x_3)\land(\lnot x_1\lor \lnot x_2\lor \lnot x_3)$.}
		\label{fig:OSAT_I_PSPACE_HARDNESS_PROOF}
	\end{wrapfigure}	
	
Finally we need to ensure that all concepts $C_j$ are true in the leaves depending on the quantifications $\Game_1 x_1\Game_2 x_2\cdots\Game_n x_n$. For this purpose, we add the following axioms to the TBox $\calT$ which ensure that, starting at the root, we run through each variable level of the tree as required by the quantification in $\varphi$, and reach only leaves that are no instances of $f$, \ie, that are instances of $f'$:
\begin{align}
	\{d_0\dsub\Game_1 R_1.\Game_2 R_2.\cdots\Game_n R_n.f'\}\label{eq:quantifiers}
\end{align}

\begin{Claim}
	$\varphi\equiv\true$ iff $\calT\in\ALCTSAT(\{\true,\false\})$.
\end{Claim}
\begin{Proofofclaim}

	``$\Leftarrow$'': Let $\calI=(\Delta^\calI,\cdot^\calI)$ be an interpretation \st $\calI\models\calT$. 
        Due to axiom \axref{eq:init}, there exists an individual $w_0$ that is an instance of $d_0$.
Because of axioms \axref{eq:varBranch} and \axref{eq:disjunctLevels}, there are at least two different $R_1$-successors of $d_0$, one being an instance of $x_1$ and the other of $x_1'$ (axiom \axref{eq:varCarry} in combination with axiom \axref{eq:disjunctLevels} ensure that these successors are fresh individuals). \emph{Every} other $R_1$-successor is an instance of either $x_1$ or $x_1'$, due to axiom \axref{eq:varNeg}. Other possible $R_j$-edges for $2\leq j\leq n$ will not affect our argumentation as we will see in the following.

Repeated application of axioms \axref{eq:varBranch} and \axref{eq:disjunctLevels} shows that this structure becomes a complete binary tree of depth $n$ with (at least) $2^n$ leaves. Each leaf represents one of all possible Boolean combinations of $x_i$ and $x_i'$ for $1\leq i \leq n$. Due to axioms \axref{eq:varBranch} and \axref{eq:disjunctLevels}, every possible combination does occur. In addition, axiom \axref{eq:clauseNeg} and \axref{eq:labelF} ensure the following: each leaf is an instance of either $C_j$ or $C_j'$, for each $1\leq j \leq m$; if a leaf is an instance of at least one such $C_j$, it is also an instance of $f$.

	Axiom \axref{eq:quantifiers} allows us to conclude that all relevant leaves that represent the assignments $\theta_i\colon\{x_1,\dots,x_n\}\to\{0,1\}$ for which $\theta_i\models C_1\land\cdots\land C_m$ must hold, are instances of the proposition $f'$. Additional $R_j$-edges, as mentioned above, do not contradict the argumentation. Hence every relevant leaf must be an instance of every $C_j$ because otherwise it were an instance of $C_j'$ and thus of $f'$. Therefore, at least one literal in each clause is labeled and thereby satisfied. Hence $\varphi\equiv\true$.

	Note that only those leaves that correspond to an assignment satisfying $C_j$ can be instances of $C_j$. To clarify this fact, consider a clause $C_j= l_{1j}\lor l_{2j}\lor l_{3j}$ that is not satisfied by some assignment $\theta\colon\{x_1,\dots,x_n\}\to\{\true,\false\}$, 
        and some leaf $w$ is (erroneously) an instance of $C_j$. As $\theta\not\models C_j$, it holds that $\theta\not\models l_{ij}$ for $1\leq i\leq 3$. Thus $l_{ij}'$ must be labeled in $w$ in order for axiom \axref{eq:varNeg} to be satisfied. Now axiom \axref{eq:invokeClause} enforces $R_{1j}$- and $R_{2j}$-edges to successors satisfying $\widetilde l_{2j}$ and $\widetilde l_{3j}$. Finally, these propositions and transitions lead to
        $w$ being an instance of $C_j'$. This is not possible because $C_j$ and $C_j'$ are disjoint due to axiom \axref{eq:clauseNeg}.

\newcommand{\calIchi}{{\calI_{\chi_1}}}
\newcommand{\calTchi}{{\calT_{\chi_1}}}
\newcommand{\calIphi}{{\calI_{\varphi}}}
\newcommand{\calTphi}{{\calT_{\varphi}}}

	``$\Rightarrow$'': Let $n$ be the number of variables in $\varphi$. 
        In \ifreport the following \else \cite{mesc10} \fi
        we show by induction on $n$:
        %TS: Ab hier habe ich einige ``\equiv'' durch ``='' ersetzt, wo ``='' gemeint war.
        if $\varphi=\exists x_1\forall x_2\cdots\Game x_n(C_1\land\cdots\land C_m)\equiv\true$, then $\calT\in\ALCTSAT(\{\true,\false\})$.

\ifreport
\emph{Induction basis.} $n=1$. \Wlog, we assume that $\varphi$ starts with $\exists$, \ie, $\varphi=\exists x_1(C_1\land\cdots\land C_m)\equiv\true$,
and we assume that each $C_i$ contains the positive literal $x_1$.

We construct a model $\calI=(\Delta^{\calI},\cdot^\calI)$ where we set $\Delta^{\calI} = \{w_0,w_1,w_2,w_3\}$, $(S)^\calI=\{(w_0,w_1),(w_1,w_1),$ $(w_2,w_1),(w_3,w_1)\}$, $(R_{1})^\calI=\{(w_1,w_2),(w_1,w_3)\}$, $(R_{x_1}\!)^\calI=\{(w_0,w_0),(w_1,$ $w_1),(w_2,w_2)\}$, $(d_0)^\calI=\{w_1\}$, $(d_1)^\calI=\{w_2,$ $w_3\}$, $(x_1)^\calI=\{w_0,w_1,w_2\}$, $(x_1')^\calI=\{w_3\}$. Then $(C_j)^\calI=\{w_0,w_1,w_2\}$, $(C_j')^\calI=\{w_3\}$ and $(R_{C_j})^\calI=\{(w_2,w_2)\}$ for all $1\leq j\leq m$. Finally $(f')^\calI=\{w_2\}$ (the remaining labels are irrelevant).
From this it can be easily verified that $\calI\models\calT$.

\emph{Induction step.} Assume it holds for $n\ge1$. In the following we will show that the proposition holds for $\varphi=\exists x_1\forall x_2\cdots \Game x_{n+1}F\equiv\true$ with $F=(C_1\land\cdots\land C_m)$. If $\varphi$ starts with $\forall$, the argumentation is analogous. %TS: Das finde ich klarer.
Let $F[x_1/\true]$ (or $F[x_1/\false]$) denote the matrix $F$ of $\varphi$ with every occurrence of $x_1$ replaced by $\true$ (or $\false$).
Since $\varphi \equiv \true$, we have that $\chi_1=\forall x_2\exists x_3\cdots\Game x_{n+1} F[x_1/\true]\equiv\true$ or $\chi_2=\forall x_2\exists x_3\cdots\Game x_{n+1} F[x_1/\false]\equiv\true$.

Assume $\chi_1=\forall x_2\exists x_3\cdots\Game x_{n+1} F[x_1/\true]\equiv\true$. Let $\calTchi$ be the consistent terminology that is constructed out of $\chi_1$ (this follows from our induction hypothesis). For each variable $x_i$ let $\calC_{i}=\set{C_j}{x_i\in C_j, 1\leq j\leq m}$ and $\calC'_{i}=\set{C_j}{\overline x_i\in C_j, 1\leq j\leq m}$ be the set of clauses that include the literal $x_i$ resp. $\overline x_i$. Let $\theta_1,\theta_2,\dots,\theta_{k}$ with $\theta_i\colon\{x_2,x_3,\dots,x_{n+1}\}\to\{0,1\}$ be the satisfying assignments generated by $\exists x_2\forall x_3\cdots\Game x_{n+1}$.
As $\calTchi$ is consistent, let $\calIchi=(\Delta^\calIchi,\cdot^\calIchi)$ be an interpretation \st $\calIchi\models\calTchi$. Hence it satisfies every axiom in $\calTchi$ (that has the form of above) and in particular axiom \axref{eq:init}. Thus there is an individual $w_0\in\Delta^\calIchi$ \st $w_0 \in(d_0)^\calIchi$ and therefore a binary tree is generated starting in $w_0$ (with the same argumentation as before). As that tree is defined over the variables $x_2,x_3,\dots,x_{n+1}$, all leaves addressed by axiom \axref{eq:quantifiers} include $C_1,C_2,\dots,C_m,f'$, and for the transition relation $R_{2}$ it holds $(R_{2})^\calIchi\supseteq\{(w_0,w_2),(w_0,w_2')\}$ with $(x_2)^\calIchi\ni w_2$, $(x_2')^\calIchi\ni w_2'$.

In the next steps we will construct an interpretation $\calIphi$ \st $\calIphi\models\calTphi$. Therefore we start with the previous interpretation $\calIchi$ and modify it into $\calIphi$ in the following steps as visualized in Figure \ref{fig:pspace-hardness-ALCOSAT_I_correctness}:

\begin{wrapfigure}[21]{r}{0.4\textwidth}
	\centering
 \includegraphics{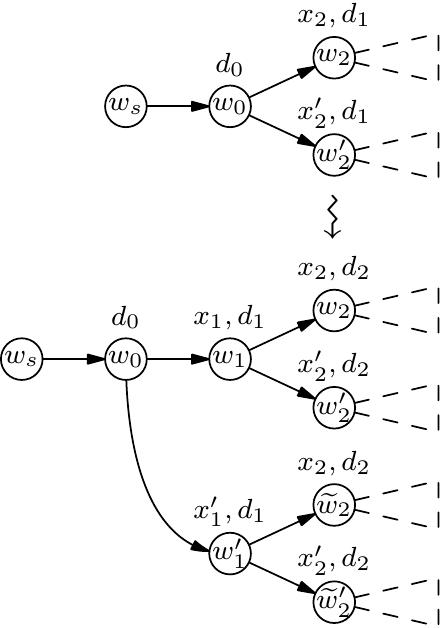}
	\caption{Construction from $\calTchi$ to $\calTphi$ in the proof of Lemma  \ref{lem:ALCOSAT_I_PSPACE_HARD}.}
	\label{fig:pspace-hardness-ALCOSAT_I_correctness}
\end{wrapfigure}
Set $\Delta^\calIphi:=\Delta^\calIchi$. Now Define $(R_{1})^\calIphi=\{(w_0,w_1)\}$ with $w_1$ a new individual added to $\Delta^\calIphi$. This inserts the first branch at the top for variable $x_1$, and hence we add $w_1$ to $(x_1)^\calI$. Now replace $(w_0,w_2),(w_0,w_2')$ from $(R_{2})^\calIphi$ with $(w_1,w_2),(w_1,w_2')$ to mount the binary tree in $\calIchi$ below the node $w_1\in\Delta^\calIphi$ that represents choosing $x_1$. Also add $w_1$ to the set $(d_1)^\calI$. Now let $\calT\subset\Delta^\calIchi$ be the set of individuals that form the binary tree in $\calIchi$ starting in $d_0$. For each individual in $\calT\setminus\{w_0\}$ we need to increment the value of the $d_i$s by one.
	Set $\widetilde\calT:=\{\widetilde w\mid w\in\calT\setminus\{w_0,w_s\}$ plus the changes on the $d_i$s as before$\}$. Add each $w\in\widetilde\calT$ to $\Delta^\calIphi$ and for each relation $R\in\ROLE\setminus\{S\}$ and each $wRw'$ with $w,w'\in\calT$ add $(\widetilde w,\widetilde w')$ to $R^\calIphi$. For each $w\in\widetilde\calT$ and each proposition $p$ with $w\in (p)^\calIchi$ also add $\widetilde w$. Now we have a complete copy (all nodes and edges) of the binary assignment tree constructed and added to our interpretation $\calIphi$.
	In the next step we mount the previously added tree below $w_0$. Here we add a new individual $w_1'$ to $\Delta^\calI$, $w_1'\in(x_1')^\calI$, $w_1'\in(d_1)^\calI$ and $(w_0,w_1')\in(R_1)^\calI$. Now we add with $(w_1',\widetilde w_2),(w_1',\widetilde w_2')\in(R_2)^\calI$ the remaining $R_2$-edges.
In the second last step we need to add all clauses (resp., clause propositions) to the tree that are satisfied by either $x_1$ or $\overline x_1$. Therefore let $R^*$ be the transitive closure of all $R_{i}$ for $1\leq i \leq n$. For all $w\in\Delta^\calIphi$ \st $w_1R^*w$ add $w$ to $(x_1)^\calIphi$ and the same for all $w_1'R^*w$ add $w$ to $(x_1')^\calIphi$. Analogously add the propositions in $\calC_1=\set{C_j}{x_1\in C_j}$ resp. $\calC_1'=\set{C_j}{\overline x_1\in C_j}$ in the same way to the individuals $w\in(x_1)^\calIphi$ resp. $w\in(x_1')^\calIphi$ and construct the respecting clause-structures around the individuals induced by axioms \axref{eq:invokeClause}.
 For each $w\in\Delta^\calI$ add a transition $(w,w_0)$ to $(S)^\calIphi$. Finally, for each individual $w\in\{w_0,w_1,w_1',w_s\}$ and each variable $x_i$ we need to add w either to $(x_i)^\calI$ or $(x_i')^\calI$. This can be done arbitrarily and is just for satisfying the axioms \axref{eq:varNeg}. Also we need to built arround those states the needed nodes and edges induced by the clause axioms \axref{eq:invokeClause}--\axref{eq:clauseNeg}.

As $\chi_1\equiv\forall x_2\cdots\Game x_n F[x_1/\true]\equiv \true$, the ``upper'' subtree starting below $w_1$ is consistent with $x_1\dsub\forall R_2.\exists R_3.\cdots \Game_{n+1} R_{n+1}.f'$ and now it follows from the hypothesis and construction that $\calIphi\models\calTphi$. This proof generalizes to arbitrary quantification blocks $\Game_1x_1\cdots\Game_nx_n$ with $\Game_i\in\{\exists,\forall\}$.
\fi
\end{Proofofclaim}

As the number of axioms in $\calT$ is polynomially bounded and the terminology is consistent if and only if the quantified Boolean formula $\varphi$ is satisfiable, the lemma applies.
\end{Proof}

\begin{lemma}\label{lem:ALCOCSAT_N2_PSPACE-HARD}
 $\ALCTSAT(\{\lnot\})$ is \PSPACE-hard.
\end{lemma}
\begin{Proof}
	From Lemma \ref{lemma:topbot-always-above-neg} we can simulate $\true$ and $\false$ with fresh atomic concepts. Then the argumentation follows similarly to Lemma \ref{lem:ALCOSAT_I_PSPACE_HARD}.
\end{Proof}

\begin{lemma}\label{lem:ALCOCSAT_I0_CONP-HARD}
	$\ALCTCSAT(\{\false\})$ is \coNP-hard.
\end{lemma}
\begin{Proof}
%  \ifreport
        In contrast to Lemma \ref{lem:ALCOSAT_I_PSPACE_HARD}, the instances of $\ALCTCSAT(\CloneI_0)$ consist of a concept $C$ and a TBox $\calT\subseteq\Ax(\{\false\})$.
        Both do not contain the concept $\true$. Now we adapt the proof of Lemma  \ref{lem:ALCOSAT_I_PSPACE_HARD} to this new setting as follows:
        in all axioms containing $\true$, we replace $\true$ with a fresh atomic concept $t$.
        This is unproblematic except for axiom \axref{eq:init}, where we need to enforce $d_0$ to have an instance.
        For this purpose, we remove the axiom $\true\dsub\exists S.d_0$ from \calT and set $C = d_0$.
        Additionally, we need to adopt axiom \axref{eq:quantifiers} to $d_0\dsub\forall R_1.\forall R_2.\cdots\forall R_n.f'$ to match the desired reduction from \TAUT.
        Please note, that with this construction it is not possible to state a reduction from \QBFthreeSAT, because an interpretation where whenever we want to branch existentially, a respective individual with neither $x_i$ nor $x_i'$ labeled can be added without interfering the axioms, in particular axiom \axref{eq:varNeg}.
%  \else
%    The proof is a slight variation of the proof of Lemma \ref{lem:ALCOSAT_I_PSPACE_HARD}, which is explained
%    in \cite{mesc10}.
%  \fi
\end{Proof}
\section{Conclusion}

\ifreport
  \begin{figure}[htpb]
  \centering

\includegraphics{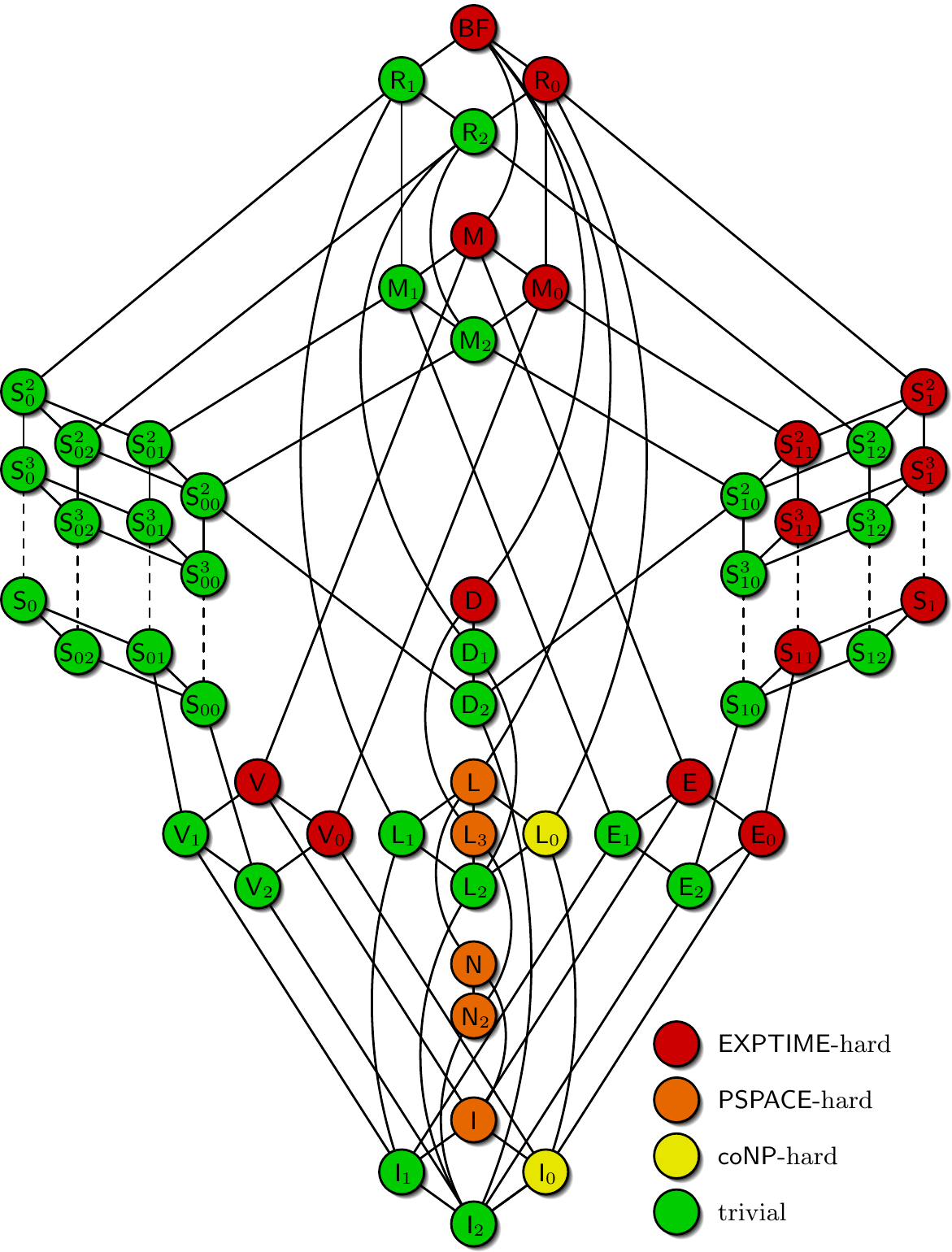}
%\begin{asy}
%     import lattice;
%     import patterns;
%     //add("hatch",hatch(H=4, dir=NE, white+3.5));
%     defaultpen(0.8+fontsize(9));
% 
%     pen EXPTIME = rgb(0.8,0,0);
%     pen PSPACEh = rgb(0.9,0.4,0);
%     pen coNPh = rgb(0.9,0.9,0);
%     pen trivial = rgb(0,0.8,0);
%     pen open = gray(1);
%   //  pen ParityL_to_P = pattern("hatch");
% 
%     Lattice lattice = Lattice(0.9cm, 0.65cm, 0.28cm);
% 
%     lattice.setUpperBound(lattice.BF,black,EXPTIME,"$\mathsf{EXPTIME}$-hard",1);
%   // lattice.setUpperBound(lattice.L,black,open, "open",11);
%     lattice.setUpperBound(lattice.R1,black,trivial, "trivial",9);
% 
%     lattice.setLowerBound(lattice.I0,black,coNPh,"$\mathsf{coNP}$-hard",3);
%     lattice.setLowerBound(lattice.I,black,PSPACEh,"$\mathsf{PSPACE}$-hard",2);
%     lattice.setLowerBound(lattice.N2,black,PSPACEh,2);
%     lattice.setLowerBound(lattice.E0,black,EXPTIME,1);
%     lattice.setLowerBound(lattice.V0,black,EXPTIME,1);
%     lattice.setLowerBound(lattice.D,black,EXPTIME,1);
% 
%     //Clone[] c = {lattice.I,lattice.BF,lattice.M,lattice.E,lattice.V,lattice.N,lattice.L};
%     //lattice.restrictTo(c);
% 
%     lattice.draw();
%         lattice.legend(point(S)+(2.5cm,0.5cm));
%\end{asy}
% TS: Wenn Leerzeichen oder tabs vor \begin{asy} oder \end{asy} stehen, gibt's komische Fehler.
% Siehe auch http://newsgroups.derkeiler.com/pdf/Archive/De/de.comp.text.tex/2006-12/msg00055.pdf

  \caption{Complexity for $\ALCTCSAT(B)$, $\ALCOSAT(B)$ and $\ALCOCSAT(B)$.}
  \label{fig:latticeALCOCSAT}

  \end{figure}
  \begin{figure}[htpb]
  \centering
  
\includegraphics{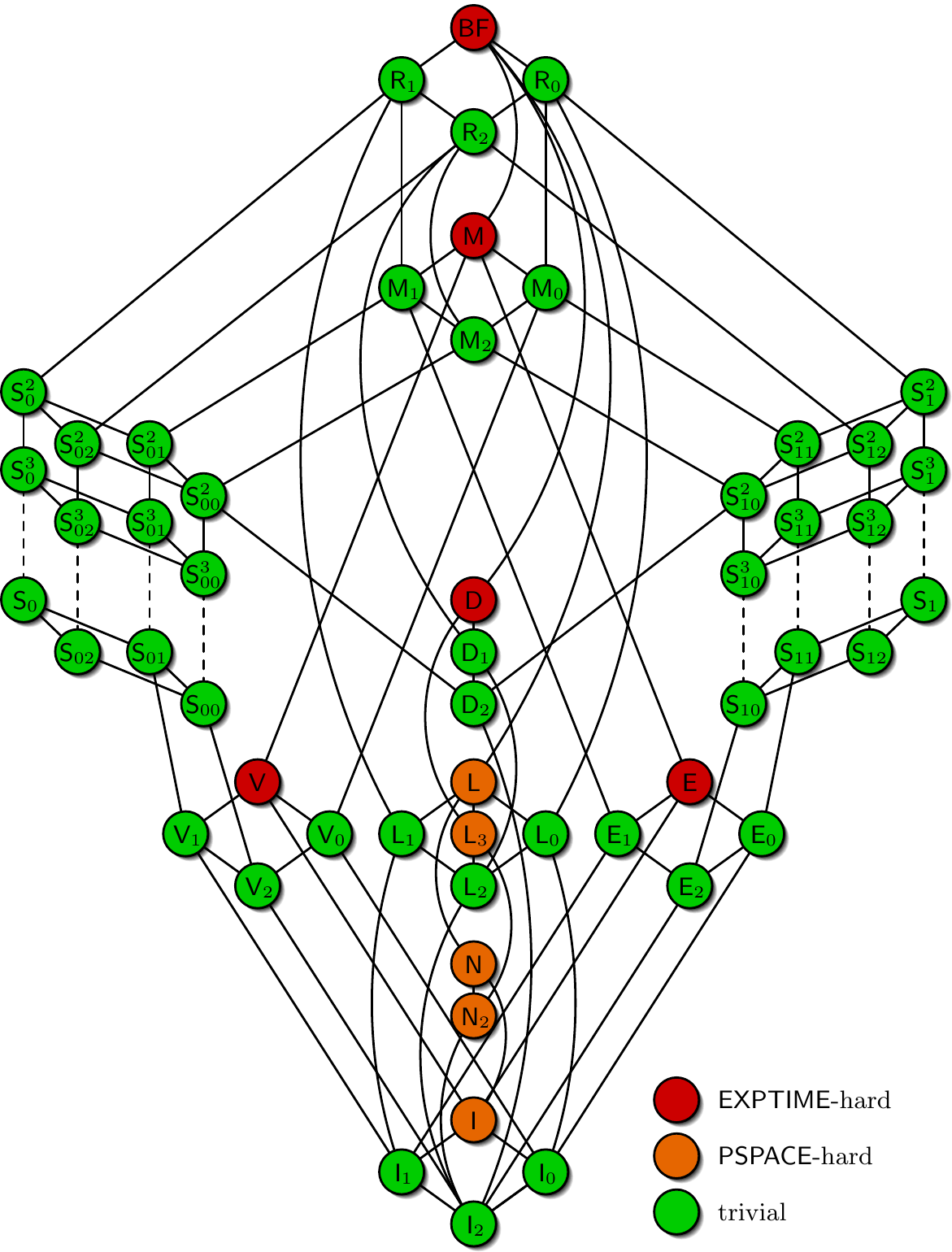}
%\begin{asy}
%     import lattice;
%     import patterns;
%     defaultpen(0.8+fontsize(9));
% 
%     pen EXPTIME = rgb(0.8,0,0);
%     pen PSPACEh = rgb(0.9,0.4,0);
%     pen coNPh = rgb(0.9,0.9,0);
%     pen trivial = rgb(0,0.8,0);
%     pen open = gray(1);
% 
%     Lattice lattice = Lattice(0.9cm, 0.65cm, 0.28cm);
% 
%     lattice.setUpperBound(lattice.BF,black,EXPTIME,"$\mathsf{EXPTIME}$-hard",1);
%     lattice.setUpperBound(lattice.R1,black,trivial, "trivial",9);
%     lattice.setUpperBound(lattice.R0,black,trivial,9);
%
%     lattice.setLowerBound(lattice.I,black,PSPACEh,"$\mathsf{PSPACE}$-hard",2);
%     lattice.setLowerBound(lattice.N2,black,PSPACEh,2);
%     lattice.setLowerBound(lattice.E,black,EXPTIME,1);
%     lattice.setLowerBound(lattice.V,black,EXPTIME,1);
%     lattice.setLowerBound(lattice.D,black,EXPTIME,1);
% 
%     lattice.draw();
%         lattice.legend(point(S)+(2.5cm,0.5cm));
%\end{asy}
% TS: Wenn Leerzeichen oder tabs vor \begin{asy} oder \end{asy} stehen, gibt's komische Fehler.
% Siehe auch http://newsgroups.derkeiler.com/pdf/Archive/De/de.comp.text.tex/2006-12/msg00055.pdf

  \caption{Complexity for $\ALCTSAT(B)$.}
  \label{fig:latticeTSAT}

  \end{figure}
\fi
With Corollaries \ref{cor:ALCOCSAT_results_full} and \ref{cor:ALCTSAT_results_full},
we have separated the problems \TSAT, \TCSAT, \OSAT\ and \OCSAT\ for \ALC fragments
obtained by arbitrary sets of Boolean operators into tractable and intractable cases.
  We have shown that these problems are on the one hand for \TSAT
  \begin{itemize}
    \item \EXPTIME-hard whenever we allow the constant $\true$ in combination with at least conjunction or disjunction,
    	\item \EXPTIME-hard whenever all Boolean self-dual functions can be expressed,
    \item \PSPACE-hard whenever we allow at least negation or both constants $\false,\true$,
    \item trivial in all other cases.
  \end{itemize}
On the other hand for the remaining three satisfiability problems we reached \EXPTIME-hardness even for only disjunction or conjunction (without the constant \true), and got \coNP-hard cases whenever we allow at least the constant $\false$ (hence the \false-reproducing cases that are trivial for \TSAT\ drop to intractable for these problems).
  
According to the
\ifreport
  Figures \ref{fig:latticeALCOCSAT} and \ref{fig:latticeTSAT},
\else
  Figures 4 and 5 in \cite{mesc10}, \fi
which arrange our results in Post's lattice,
this classification covers %almost 
all sets of Boolean operators
closed under nesting. 
%The only open cases are caused by a currently missing $\coNP$-hardness result for $\TSAT(B)$ with $[B]\subseteq\CloneI_0$. The main reason is that we cannot express $\top$, and therefore we cannot enforce the start of the assignment tree.

We have also shown how our results, and the direct transfer of the
results in \cite{hescsc08} to concept satisfiability, generalise known results for the \FL and \EL family and other fragments of \ALC. Furthermore, due to the presence of arbitrary axioms,
the overall picture differs from similar analyses
for related logics in that hardness already holds for small sets of inexpressive Boolean operators.

It remains for future work to find matching upper bounds for the hardness
results, to look at fragments with only existential or universal restrictions, and to
restrict the background theories to terminologies with atomic left-hand sides of
concept inclusion axioms with and without cycles. Furthermore, since the standard
reasoning tasks are not always interreducible if the set of Boolean operators is restricted,
a similar classification for other decision problems such as concept subsumption is pending.

\subsection*{Acknowledgements}
We thank Peter Lohmann and the anonymous referees for helpful comments and suggestions.

%% Bibliography
% \bibliographystyle{splncs}
\bibliographystyle{plain}
\bibliography{description_logic}

\end{document}